# A Simple Analytical Model of Coupled Single Flow Channel over Porous Electrode in Vanadium Redox Flow Battery with Serpentine Flow Channel

Xinyou Ke[a, b, e, *], J. Iwan D. Alexander[a, c, d], Joseph M. Prahl[a], Robert F. Savinell[b, c]

[a]Fluid Mechanics Lab, Department of Mechanical and Aerospace Engineering, Case Western Reserve University, Cleveland, Ohio 44106, United States

[b]Electrochemical Engineering and Energy Lab, Department of Chemical Engineering, Case Western Reserve University, Cleveland, Ohio 44106, United States

[c]Great Lakes Energy Institute, Case Western Reserve University, Cleveland, Ohio 44106, United States

[d]Current address: School of Engineering, University of Alabama at Birmingham, Birmingham, Alabama 35294, United States

[e]Current address: School of Engineering and Applied Science, Harvard University, Cambridge, Massachusetts 02138, United States

[*]Corresponding author: Xinyou Ke; Email address: xinyouke@gmail.com

**Abstract**

A simple analytical model of a layered system comprised of a single passage of a serpentine flow channel and a parallel underlying porous electrode (or porous layer) is proposed. This analytical model is derived from Navier-Stokes motion in the flow channel and Darcy-Brinkman model in the porous layer. The continuities of flow velocity and normal stress are applied at the interface between the flow channel and the porous layer. The effects of the inlet volumetric flow rate, thickness of the flow channel and thickness of a typical carbon fiber paper porous layer on the volumetric flow rate within this porous layer are studied. The maximum current density based on the electrolyte volumetric flow rate is predicted, and found to be consistent with reported numerical simulation. It is found that, for a mean inlet flow velocity of 33.3 cm s$^{-1}$, the analytical maximum current density is estimated to be 377 mA cm$^{-2}$, which compares favorably with experimental result reported by others of ~400 mA cm$^{-2}$.

# 1. Introduction

Studies on redox flow batteries (RFBs) with flow fields (such as serpentine flow channels and interdigitated flow channels) that evolved from PEM fuel cell designs have gained more attention. The thin carbon fiber paper electrodes used in the RFBs with flow fields give much lower ohmic losses compared with the conventional flow batteries with thicker carbon felt electrodes. A much higher current density can be achieved in the RFBs with the thin carbon fiber electrodes.

Results reported by Zawodzinski and Mench et al. [1] of the "zero-gap" flow battery prototype design utilizing an architecture of the serpentine flow channel flow field demonstrate performance at high current density and power density in contrast to flow batteries without flow fields. Xu et al. [2] compared cell performance of the vanadium RFBs between without and with flow fields. Higher round trip efficiency is found in the latter. Latha et al. [3] studied the flow dynamics, including flow velocities and pressure drops in the all-vanadium RFBs with the serpentine flow channels. Ke et al.'s [4, 5] macroscopic mathematical model of mass transport in RFBs with a single passage of a serpentine flow channel shows flow distributions in the porous electrode, flow reactants consumption within the porous layer and predicts a maximum current density in agreement with experimental data reported in the literature. Brian et al. [6] developed an organic Quinone-Bromine flow battery with interdigitated flow channels, and high power density has been observed in this type of flow cell. Mayrhuber et al. [7] used a laser perforation approach to generate holes among the structures of the carbon fiber paper electrodes in the RFBs with the serpentine flow channels. It has been found that those holes can help to enhance electrolyte flow and consequently mass transfer and penetration in the porous electrode and a higher power density can be achieved. Nevertheless, excessive holes can result in smaller active surface area, which leads to a degradation of the performance.

In this work, a simple analytical model is presented to help understand high performance achieved in the vanadium flow battery with the serpentine flow channels. The model can be applied to any flow battery half-cell that has a true redox reaction with all species soluble in the electrolyte and where a serpentine flow field is used for enhanced power density. These include the all vanadium and iron-vanadium systems, the iron-chrome systems and the positive half-cells of the all iron and metal-halogen systems.

## 2. Zero-gap flow cell construction

This "zero-gap" flow cell architecture shown as Fig. 1 has been first reported by Zawodzinski and Mench et al. [1] in the flow batteries with thin carbon fiber paper electrode, which gives a much smaller ohmic loss while enabling higher limiting current density. This flow cell is modified from the fuel cell configuration of a direct methanol fuel cell (DMFC) or a proton exchange membrane fuel cell (PEMFC) design.

## 3. 2D Flow structure

The 2D *(X, Y)* diagram of the electrolyte flowing through a single passage of the serpentine flow channel and over the porous electrode is described in Fig. 2. The parameters $t_f$, $t_p$ and *L* are the thickness of the flow channel, thickness of the porous layer and length of the flow channel/porous layer, respectively. Three boundaries are denoted by *BC*#1 (interface boundary between the current collector and the flow channel), *BC*#2 (interface boundary between the flow channel and the porous layer) and *BC*#3 (interface boundary between the porous layer and the ion selective membrane).

## 4. Analytical model

This macroscopic model for the flow motions in the single passage of the serpentine flow channel and the porous layer was proposed and examined numerically by Ke et al. [4, 5]. The assumptions of the fluid flow model include incompressible and Newtonian fluid, steady state and laminar flow with no gravity effect. The results reported in ref. [4] illustrates the flow physics in the flow channel and the porous layer along both the *X* and *Y* directions. Further details of deriving the motion equations can be found in the ref. [5]. The flow patterns evolving from an entrance profile to developing, developed and fully developed regime in the flow channel and the porous layer under the ideal plug flow and parabolic flow inlet boundary conditions were studied.

In this contribution, a simple analytical solution of the model proposed and numerically analyzed by Ke et al [4, 5] is presented. As fully developed region in the porous layer is approached, the velocities in the *Y* direction approach to be zero. Under this condition, the total volumetric flow rate transferred from the flow channel into the porous layer is estimated. The flow motion along the *X* direction in the porous layer can be written as

$$\varepsilon \nabla \langle p_p \rangle = \mu \nabla^2 \langle u_p \rangle - \frac{\mu \varepsilon}{k} \langle u_p \rangle \tag{1}$$

In the flow channel, the Navier-Stokes motion is simplified as

$$0 = -\nabla p_f + \mu \nabla^2 u_f \tag{2}$$

In the fully developed region, the pressure gradient along the flow channel and the porous layer is equal. The boundary conditions are defined as follows

(a) *BC*#1-upper wall of the flow channel: $0 \leq X \leq L$, $Y=t_f$, $u_f=0$;

(b) *BC*#2-interface between the flow channel and the porous layer:

$0 \leq X \leq L$, $Y=0$, $u_f=<u_p>$; $\partial u_f/\partial Y = \varepsilon^{-1} \partial <u_p>/\partial Y$ (continuities of flow velocity and normal stress at the interface) [5];

(c) BC#3-bottom wall of the porous layer: $0 \leq X \leq L$, $Y=-t_p$, $<u_p>=0$.

As the fully developed region is approached, the conservation of volumetric flow rate gives

$$Q_{in} = (Q_f)_{fd} + (Q_p)_{fd} \tag{3}$$

Where, the total of volumetric flow rate $Q_{in}$ is equal to the volumetric flow rates in the channel plus the porous layer, $(Q_f)_{fd}+(Q_p)_{fd}$. Where, $Q_{in}$, $(Q_f)_{fd}$ and $(Q_p)_{fd}$ are denoted as follows

$$Q_{in} = u_{in} t_f w_f \tag{4}$$

$$(Q_f)_{fd} = w_f \int_0^{t_f} (u_f)_{fd} \, dY \tag{5}$$

$$(Q_p)_{fd} = w_p \int_{-t_p}^{0} (u_p)_{fd} \, dY \tag{6}$$

Here, $w_f$ is equal to $w_p$. Through applying the conditions of *BC*#1, *BC*#2, *BC*#3 and conservation of volumetric flow rate, the analytical solutions for velocities and volumetric flow rates in the fully developed region of the flow channel and the porous layer are obtained, respectively

$$(u_f)_{fd} = 0.5 \frac{C_0}{\mu} Y^2 + C_1 Y + C_2 \tag{7}$$

$$(\langle u_p \rangle)_{fd} = C_3 e^{\sqrt{\frac{\varepsilon}{k}}Y} + C_4 e^{-\sqrt{\frac{\varepsilon}{k}}Y} + C_5 \tag{8}$$

The corresponding equations of volumetric flow rate in the flow channel and the porous layer are shown below

$$(Q_f)_{fd} = \frac{1}{6}\frac{C_0}{\mu} t_f^3 w_f + 0.5 C_1 t_f w_f + C_2 t_f w_f \tag{9}$$

$$(Q_p)_{fd} = \sqrt{\frac{k}{\varepsilon}} w_p \left( C_3 \left(1 - e^{-\sqrt{\frac{\varepsilon}{k}}t_p}\right) - C_4 \left(1 - e^{\sqrt{\frac{\varepsilon}{k}}t_p}\right) \right) + C_5 t_p w_p \tag{10}$$

Here,

$$C_0 = \frac{u_{in} t_f \alpha_1}{\alpha_1 \alpha_2 - \alpha_3 \alpha_4} \tag{11}$$

$$C_1 = \frac{(C_3 - C_4)}{\varepsilon} \sqrt{\frac{\varepsilon}{k}} \tag{12}$$

$$C_2 = C_3 + C_4 - \frac{C_0 k}{\mu} \tag{13}$$

$$C_3 = \frac{C_0 k}{\mu} e^{\sqrt{\frac{\varepsilon}{k}}t_p} - C_4 e^{2\sqrt{\frac{\varepsilon}{k}}t_p} \tag{14}$$

$$C_4 = \frac{\alpha_3}{\alpha_1} C_0 \tag{15}$$

$$C_5 = -\frac{C_0 k}{\mu} \tag{16}$$

Where,

$$\alpha_1 = \frac{t_f}{\varepsilon}\sqrt{\frac{\varepsilon}{k}}\left(e^{2\sqrt{\frac{\varepsilon}{k}}t_p} + 1\right) + e^{2\sqrt{\frac{\varepsilon}{k}}t_p} - 1 \tag{17}$$

$$\alpha_2 = \frac{t_f^3}{6\mu} + e^{\sqrt{\frac{\varepsilon}{k}}t_p}\left(\frac{t_f^2 k\sqrt{\frac{\varepsilon}{k}}}{2\varepsilon\mu} + \frac{kt_f}{\mu} + \frac{k}{\mu}\sqrt{\frac{k}{\varepsilon}}\right) - \frac{k}{\mu}\left(t_f + t_p + \sqrt{\frac{k}{\varepsilon}}\right) \tag{18}$$

$$\alpha_3 = \frac{t_f^2}{2\mu} + e^{\sqrt{\frac{\varepsilon}{k}}t_p}\frac{k}{\mu}\left(\frac{t_f}{\varepsilon}\sqrt{\frac{\varepsilon}{k}} + 1\right) - \frac{k}{\mu} \tag{19}$$

$$\alpha_4 = \left(e^{2\sqrt{\frac{\varepsilon}{k}}t_p} + 1\right)\left(\frac{t_f^2}{2\varepsilon}\sqrt{\frac{\varepsilon}{k}} + \sqrt{\frac{k}{\varepsilon}}\right) + \left(e^{2\sqrt{\frac{\varepsilon}{k}}t_p} - 1\right)t_f - 2e^{\sqrt{\frac{\varepsilon}{k}}t_p}\sqrt{\frac{k}{\varepsilon}} \tag{20}$$

$C_0$ is a negative value and it represents the pressure gradient along the flow channel and the porous layer. The local flow velocity and volumetric flow rates are a function of $t_f$, $t_p$, $w_f$, $w_p$, $k$, $\varepsilon$, $\mu$ and $u_{in}$ or $Q_{in}$. $u_{in}$ is the average value of the inlet velocity or the mean inlet velocity. Eq. (7) illustrates that the maximum flow velocity in the flow channel occurs at $y=-\mu C_1/C_0$, which is always smaller than the value of $t_f/2$. $C_2$ is the flow velocity at the interface between the flow channel and the porous layer.

## 5. Parameters for calculations

The dimensions of a typical lab hardware flow cell modified from direct methanol fuel cell (DMFC) or proton exchange membrane fuel cell (PEMFC). Sorts of parameters used in the calculations are given in Table 1. The thickness of a single porous layer for the 10 AA carbon fiber paper from SGL technologies is ~0.041 cm [1] and the Toray carbon paper from ref. [8] is ~0.02 cm.

## 6. Results and discussion

*6.1 Flow distributions in the porous layer*

In the Fig. 3, the flow distribution in the porous layer is illustrated when the fully developed flow regime is assumed. As the depth of the porous layer increases, the $X$ direction velocity decreases. The volumetric flow rate in the porous layer is calculated through the integration of the $X$ direction velocity $(<u_p>)_{fd}$ along the $Y$ direction in the porous layer as shown in Fig. 3. The expression for describing the flow velocities in the porous layer is given in eq. (8).

*6.2 Volumetric flow rate in the porous layer*

The volumetric flow rate in the porous layer can be calculated by eq. (10). Fig. 4 shows relationships between the volumetric flow rates in a single layer of the 10 AA carbon fiber paper electrode under four different thicknesses of the flow channel: 0.05 cm, 0.1 cm, 0.15 cm and 0.2 cm. The mean inlet flow velocity ranges from 8.3 cm s$^{-1}$ to 50 cm s$^{-1}$. The calculations are in good agreement with numerical simulation reported earlier [4, 5]. It is found that the volumetric flow rate in the porous layer almost has a linear relation with the mean inlet flow velocity. The relationships between the volumetric flow rates in the porous layer and the inlet volumetric flow rates ranging from 0.5 cm$^3$ min$^{-1}$ to 25 cm$^3$ min$^{-1}$ under three thicknesses of the Toray carbon fiber electrode: 0.02 cm (single layer), 0.04 cm (two layers) and 0.06 cm (three layers) are shown in Fig. 5. The calculations show more volumetric flow penetration into the porous layer with a thick porous layer and a larger inlet volumetric flow rate. When the thickness of the porous layer is larger than ~0.06 cm, the analytical solutions calculated by MATLAB (2012a) in a 32 G physical memory lab computer failed. This was because when $\alpha_1\alpha_2-\alpha_3\alpha_4$ approaches to be zero, $C_0$ is very sensitive to $\alpha_1\alpha_2-\alpha_3\alpha_4$. The current lab computer's limitations wrongly calculate $\alpha_1\alpha_2-\alpha_3\alpha_4$ to be precisely zero in error. A larger memory and higher level-computing cluster are needed to find out all solutions for this analytical model in the limit of very small $\alpha_1\alpha_2-\alpha_3\alpha_4$. An analytical solution often provides the benefit that one can examine limits on performance. In this case, the various terms ($C_0$, $C_1$, $C_2$, $C_3$, etc) are interdependent of the term of $t_p\sqrt{\frac{\varepsilon}{k}}$, and clear conclusions are not obvious. But, by having the analytical solution available, one can easily program the solution to investigate various parameters within the constraints of practical designs and operating conditions.

*6.3 Maximum current density*

Ke et al. [4, 5] estimated the maximum current density in RFBs with the serpentine flow channels limited by the amount of reactant penetrating into the porous electrode

$$i_{max} = \frac{nFc(Q_p)_{fd}}{w_p L} \tag{21}$$

Here, *n* is the number of electrons transferring in the chemical reactions, *F* is Faraday constant, *c* is the bulk concentration, $(Q_p)_{fd}$ is the volumetric flow rate in the porous layer, $w_p$ (equal to $w_f$) is the width of the porous layer and *L* is the length of the flow channel or the porous layer. Tang [8] also proposed a model to calculate the theoretical maximum current density in vanadium RFBs with the serpentine flow channels as shown

$$i_{max} = \frac{nFcQ_{in}}{A} \quad (22)$$

Here, $Q_{in}$ is the inlet volumetric flow rate and $A$ is the area of the ion selective membrane. However, eq. (22) does not consider the actual amount of flow penetration into the porous layer but instead assumes that all the reactant flow into the battery is accessible for reaction. The maximum current density estimated by eq. (22) is far beyond the observed value [8, 9]. As pointed out earlier the numerical model is in agreement with the limited published data [1]. In the analytical model presented here, the maximum current density predicted by eq. (21) is 377 mA cm$^{-2}$, is compared with numerical estimate of 377 mA cm$^{-2}$ (in good agreement with the experimental value of ~400 mA cm$^{-2}$) for a single layer of the 10 AA carbon fiber paper when the mean entrance flow velocity is 33.3 cm s$^{-1}$ [4,5].

## 7. Conclusions

A simple analytical model is developed to calculate the volumetric flow rate in the porous layer of the vanadium RFBs with the serpentine flow channels. The results of the analytical model calculations compare favorably with numerical simulations reported in the literature, which have already been compared to the limited experimental data reported in the literature. The maximum current density achieved in the porous layer is a function of entrance volumetric flow rate, thickness of the flow channel, width of the flow channel, thickness of the porous layer, width of the porous layer, length of the flow channel/porous layer, electrolyte density, electrolyte viscosity, electrode porosity, permeability, ion concentration and number of electrons transferring in the reactions. This analytical model should be of help to design and optimize flow battery performance.

**Acknowledgement**

This work is partially supported by the ARPA-E project (DE-AR0000352) funded from Department of Energy (DOE) of the United States. We appreciate discussions with Dr. Nathaniel C. Hoyt and Dr. Donald L. Feke from Case Western Reserve University.

**Nomenclature**

| | |
|---|---|
| $A$ | area of ion selective membrane (cm$^2$) |
| $BC$ | boundary condition |
| $C$ | constant |
| $c$ | concentration (mol cm$^{-3}$) |
| $F$ | Faraday constant (96485 C mol$^{-1}$) |
| $i$ | current density (A cm$^{-2}$) |
| $k$ | permeability of the porous electrode (cm$^2$) |
| $L$ | length (cm) |
| $P$ | pressure (Pa) |
| $<P>$ | average pressure (Pa) |
| $Q$ | volumetric flow rate (ml min$^{-1}$ or cm$^3$ s$^{-1}$) |
| $t$ | thickness (cm) |
| $u$ | $X$ direction velocity (cm s$^{-1}$) |
| $<u>$ | average $X$ direction velocity (cm s$^{-1}$) |
| $w$ | width (cm) |
| $X$ | $X$ direction |
| *Y* | *Y* direction |

Greek symbols

| | |
|---|---|
| *ε* | porosity |
| *μ* | dynamic viscosity (Pa·s) |
| *ρ* | density of electrolyte flow (kg m$^{-3}$) |
| *Σ* | interface |
| *Ω* | domain |

Subscripts

| | |
|---|---|
| *avg* | average value |
| *cf* | between the current collector and the flow channel |
| *e* | entrance |
| *f* | flow domain |
| *fd* | fully developed region |
| *fp* | between the flow channel and the porous layer |
| *in* | inlet |
| *max* | maximum |
| *p* | porous domain |
| *pm* | between the porous layer and the ion selective membrane |

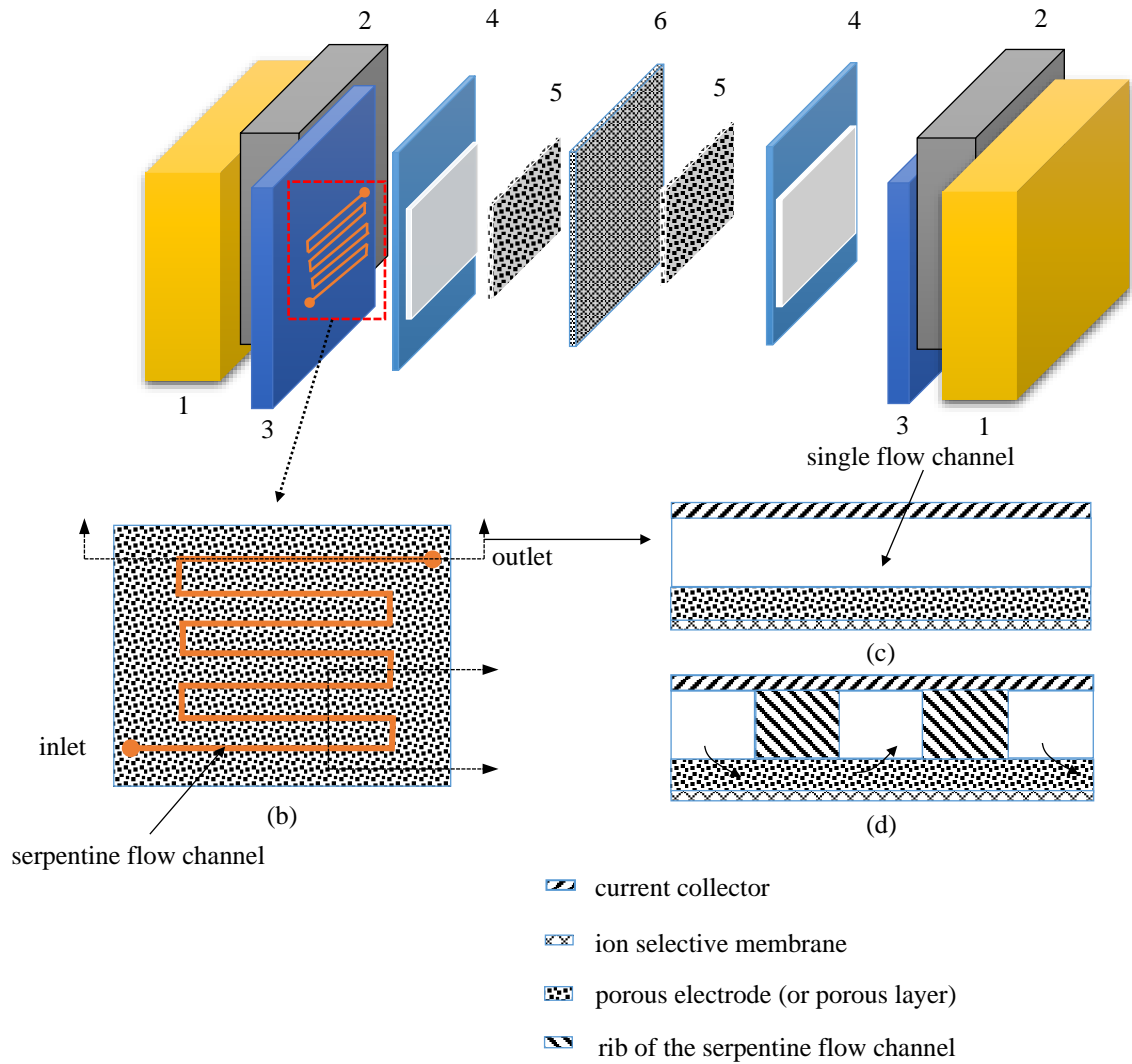

**Figure 1:** The flow cell architecture: 1-end plates, 2-current collectors, 3-graphite plates with engraved serpentine flow channels, 4-gaskets, 5-carbon fiber paper electrodes, 6-ion selective membrane; (b) serpentine flow channel with carbon fiber paper electrode (current collector and ion selective membrane are not represented); (c) a single passage of the serpentine flow channel, cross section view of (b); (d) adjoin flow channels, cross section view of (b).

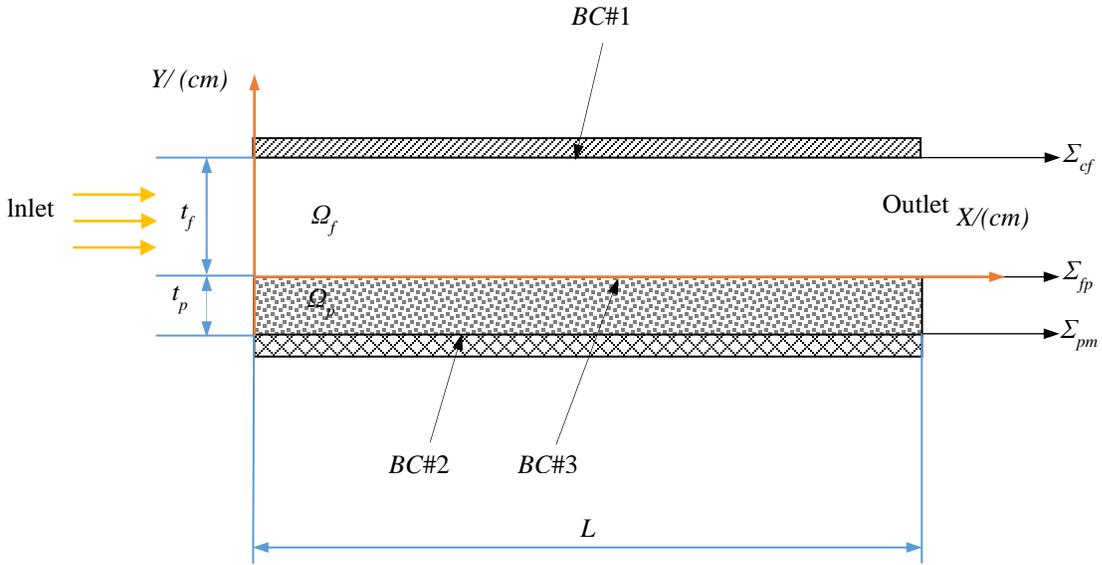

**Figure 2:** Diagram of a 2D model of flow through a single passage of the serpentine flow channel and over the under-layering porous electrode as shown in Fig. 1(c).

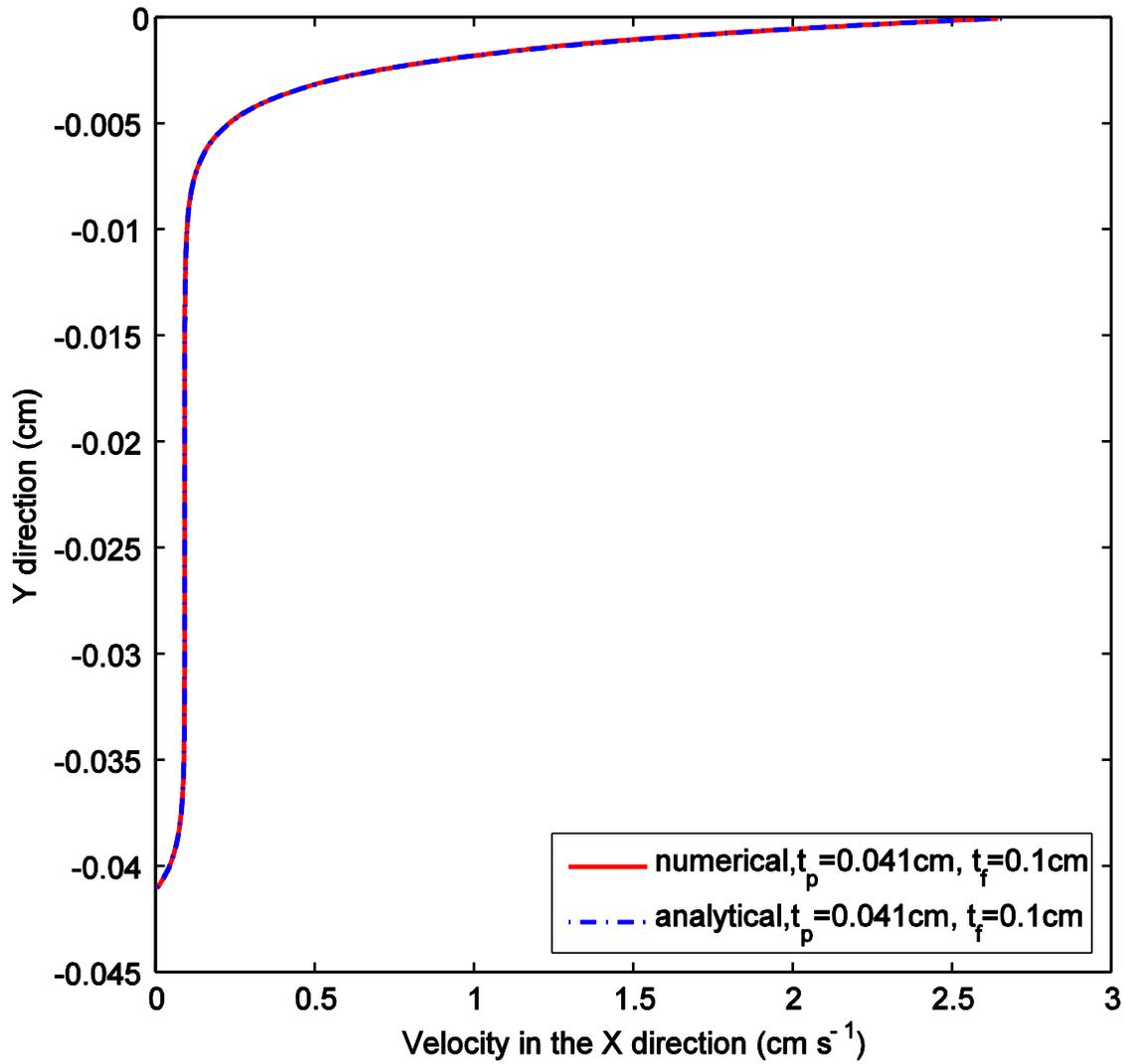

**Figure 3:** Flow velocity distributions at the fully developed region in the porous layer, $<u_p>$ (single porous layer of 10 AA carbon fiber paper, $t_f$=0.1 cm, $t_p$=0.41 cm, $u_{in}$=33.3 cm s$^{-1}$) and result of numerical simulation under the ideal plug flow inlet boundary condition.

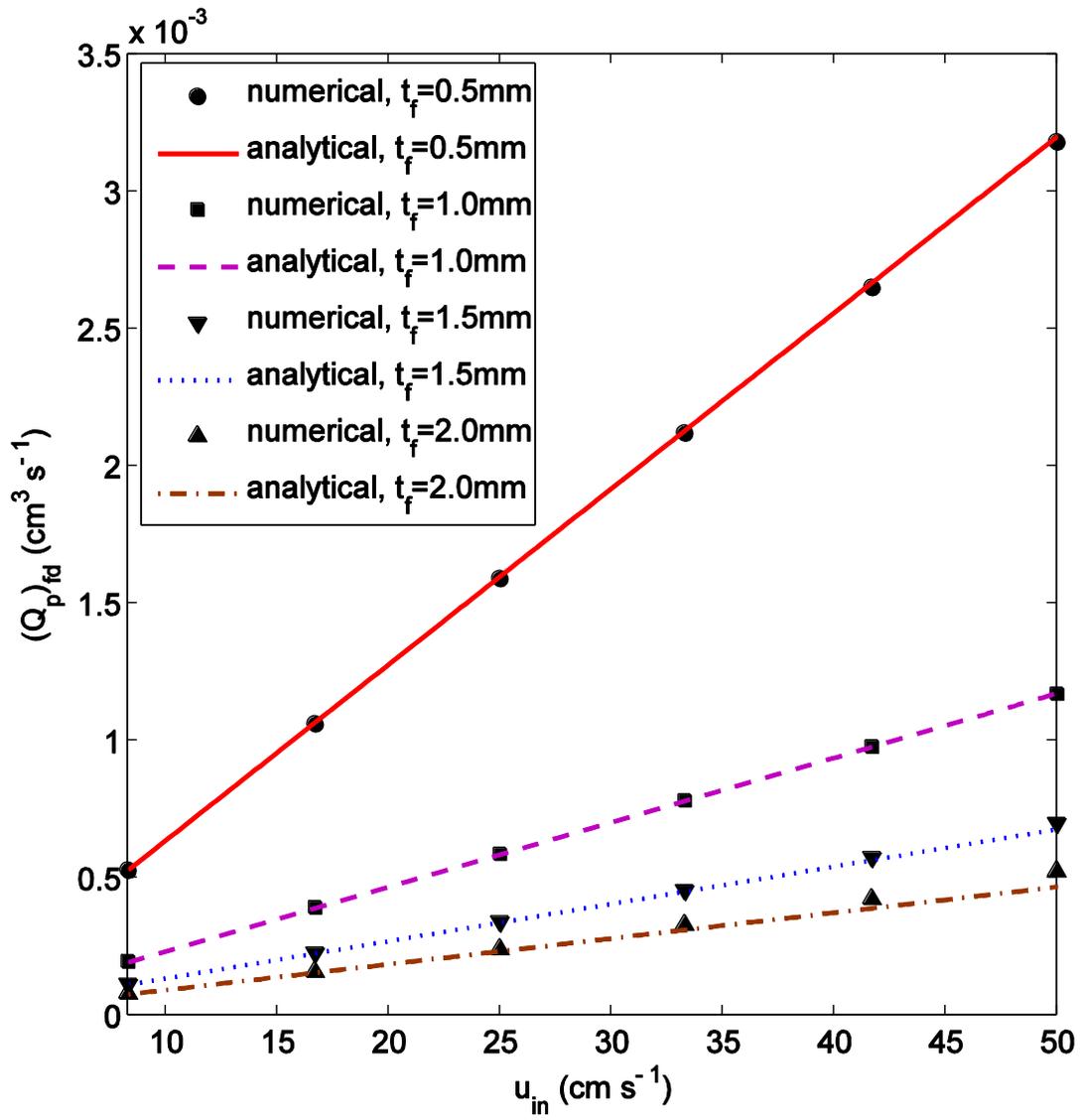

**Figure 4:** The volumetric flow rate in the porous layer $(Q_p)_{fd}$ as related to the mean inlet flow velocity $u_{in}$ for a single layer of the 10 AA carbon fiber paper electrode.

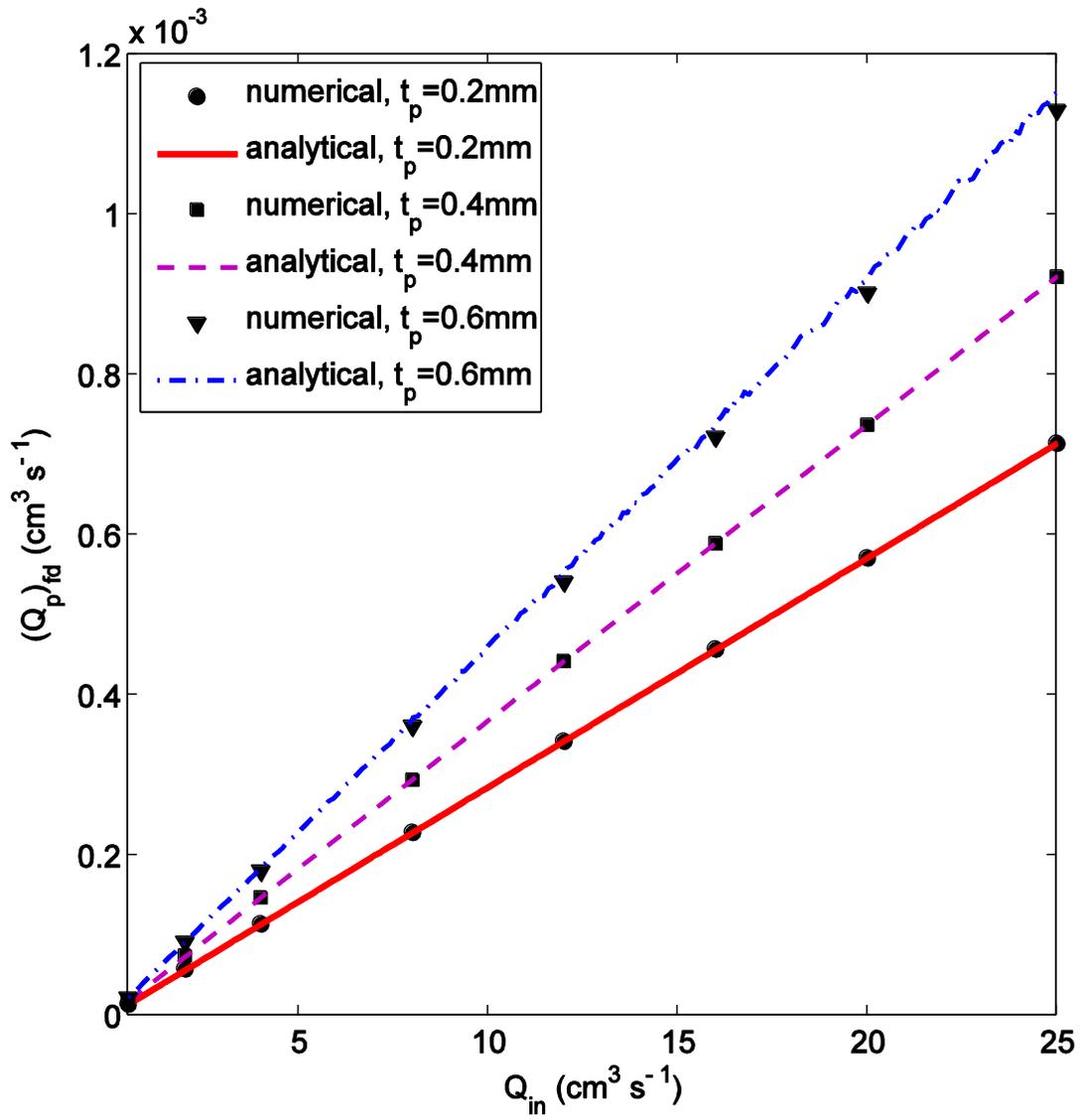

**Figure 5:** The volumetric flow rate in the porous layer $(Q_p)_{fd}$ as related to the inlet volumetric flow rate $Q_{in}$ as the number of layers for the Toray carbon fiber paper electrode increases from 1 to 3.

**Table 1** Parameters of the flow channel, porous layer, electrolyte flow and initial operation conditions

| Symbols | Descriptions | Value | Sources |
|---------|--------------|-------|---------|
| $L$ | Length of flow channel | 2 cm | |
| $w_f$ | Width of flow channel | 0.1 cm | |
| $t_f$ | Thickness of flow channel | 0.1 cm | |
| $\varepsilon$ | Porosity | 0.8 | |
| $k$ | Permeability | $2.31 \times 10^{-6}$ cm$^2$ | [4, 5] |
| $\rho$ | Density of electrolyte | 1.35 g cm$^{-3}$ | |
| $\mu$ | Dynamic viscosity of electrolyte | $4.93 \times 10^{-3}$ Pa·s | |
| $T$ | Working temperature | 298 K | |
| $c$ | Ion concentration | 0.001 mol cm$^{-3}$ | |
| $u_{in}$ | Mean inlet flow velocity | 33.3 cm s$^{-1}$ | |